\documentclass[10pt, two column, twoside]{IEEEtran}

\usepackage{amssymb}
\usepackage{amsmath}
\usepackage[lined,boxed,commentsnumbered, ruled]{algorithm2e}
\usepackage{mathrsfs}
\usepackage{algorithmic}
\usepackage{bm}

\usepackage{tikz}
\usetikzlibrary{arrows}
\usepackage{subfigure}
\usepackage{graphicx,booktabs,multirow}

\definecolor{colorhkust}{RGB}{20,43,140}
\definecolor{colortsinghua}{RGB}{116,52,129}
\definecolor{color1}{RGB}{128,0,0}

\newcommand{\trace}{{\rm Tr}}
\newcommand{\Sym}{{\mathrm{Sym}}}
\newcommand{\grad}{\mathrm{grad}}
\newcommand{\rc}{\nabla}
\newcommand{\D}{\mathrm{D}}
\newcommand{\GL}[1]{{\mathrm{GL}({#1})}}
\newcommand{\mat}[1]{{\bm #1}}
\newcommand{\rankk}{\mathrm{rank}}
\newcommand{\mini}{\sf{minimize}}
\newcommand{\maxi}{\sf{maximize}}
\newcommand{\subj}{\sf{subject~to}}
\newcommand{\hess}{\rm{Hess}}
\newcommand{\diagg}{\rm{diag}}

\newcommand{\subject}{\sf{subject~to}}

\newcommand{\changeBMM}[1]{\textcolor{black}{#1}}

\newcommand{\changeBM}[1]{#1}

\begin{document}

% \title{Scalable  Coordinated Beamforming for  Large-Scale Cooperative Wireless Networks}
% \author{Yuanming~Shi,~\IEEEmembership{Student Member,~IEEE,}
%         Jun~Zhang,~\IEEEmembership{Member,~IEEE,}
%         and~Khaled~B. Letaief,~\IEEEmembership{Fellow,~IEEE}% <-this % stops a space
% % \thanks{This work is supported by the Hong Kong Research Grant Council under
% % Grant No. 610212 and 610311. The work of J. Zhang is also supported by the
% % Hong Kong RGC Direct Allocation Grant DAG11EG03. A conference version of this paper has been presented in the IEEE Global Communications Conference (GLOBECOM), Atlanta, GA, USA, Dec. 2013.}
% \thanks{The authors are with the Department of Electronic and Computer Engineering, The Hong Kong University of Science and Technology, (e-mail: \{yshiac, eejzhang, eekhaled\}@ust.hk).}
% }

\title{Topological Interference Management with  User Admission Control via Riemannian Optimization}
\author{Yuanming~Shi,~\IEEEmembership{Member,~IEEE,}
        and Bamdev Mishra
\thanks{Y. Shi is with the School of Information Science and Technology,
ShanghaiTech University, Shanghai, China (e-mail: shiym@shanghaitech.edu.cn).}
\thanks{B. Mishra is with Amazon Development Centre India, Bangalore,
Karnataka 560055, India (e-mail: bamdevm@amazon.com).}}
                           
\maketitle
\IEEEpeerreviewmaketitle

% The paper headers
%\markboth{Journal of \LaTeX\ Class Files,~Vol.~6, No.~1, January~2007}%
%{Shell \MakeLowercase{\textit{et al.}}: Bare Demo of IEEEtran.cls for Journals}

\maketitle
%\doublespacing

\begin{abstract}
Topological interference management (TIM) provides a promising way to manage interference only based on the network connectivity information. Previous works on the TIM problem mainly focus on using the index coding approach and graph theory to establish conditions of network topologies to achieve the feasibility of topological interference management. In this paper, we propose a novel user admission control approach via sparse and low-rank optimization to maximize the number of admitted users for achieving the feasibility of topological interference management. To assist efficient algorithms design for the  formulated rank-constrained (i.e., degrees-of-freedom (DoF) allocation) $\ell_0$-norm \emph{maximization} (i.e., user capacity maximization) problem, we propose a  regularized smoothed $\ell_1$-norm minimization approach to induce sparsity pattern, thereby guiding the user selection. We further develop a  Riemannian trust-region algorithm to solve the resulting rank-constrained smooth optimization problem via exploiting the quotient manifold of fixed-rank matrices. Simulation results demonstrate the effectiveness and near-optimal
performance of the proposed Riemannian algorithm to maximize the number of admitted users for topological interference management.     
\end{abstract}

 \begin{IEEEkeywords}
Topological interference management, user admission control, sparse and low-rank modeling, Riemannian optimization, quotient manifold.  
 \end{IEEEkeywords}

\IEEEpeerreviewmaketitle

\section{Introduction}
The popularization of innovative applications and new services, such as Internet of Things (IoT)  and wearable devices \cite{Dohler_JSAC16IoT}, is driving the era of wireless big data \cite{Ding_CMagBigData}, thereby revolutionizing the segments of the society. In particular, with ultra-low latency and ultra-reliable requirements, Tactile Internet \cite{Fettweis_JSAC2016} enables a new paradigm shift from content-delivery to skill-set delivery networks. Network densification \cite{Bhushan_2014networkdensification, Andrews2015we}, supported by the advanced wireless technologies (e.g., massive MIMO \cite{Rusek_SPM2013}, Cloud-RAN \cite{Yuanming_TWC2014, Peng_HCRAN}, and small cells \cite{Tony_2013small, Jeff_JSAC5G}), becomes the key enabling technology to accommodate the exponential mobile data traffic growth, as well as provide ubiquitous connectivity for massive devices. However, by adding more radio access points per volume, interference becomes becomes the  bottleneck to harness the benefits of network densification. Although the recent development of interference alignment \cite{Jafar_IT2008} and interference coordination \cite{Gesbert_JSAC10} have been shown to be effective in the interference-limited communication scenarios, the significant signaling overhead of obtaining the global channel state information (CSI) limits applicability to dense wireless networks \cite{Jafar_TIT2013TIM}.

To reduce the CSI acquisition overhead and make it scalable in dense wireless networks, topological interference management (TIM) approach was proposed in \cite{Jafar_TIT2013TIM} to manage interference only based on the network connectivity information. However, establishing the feasibility of topological interference management is a challenging task. In the slow fading scenario, i.e., channels stay constant during transmission, the TIM problem turns out to be equivalent to the  index coding problem \cite{Kol_TIT2011index}, which is, however, NP-hard in general and only some special cases have been solved \cite{Jafar_TIT2014indexcoding, Jafar_TIT2013TIM}.  Furthermore, the topological interference management
with transmitters cooperation and multiple transmitter antennas were investigated
in \cite{Gesbert_TIT2015TIM} and \cite{Jafar_ISIT2014TIM_MIMO}, respectively. In the fast fading scenario, the graph theory and matroids theory were adopted to find the conditions of network topologies to achieve a certain amount of DoF allocation \cite{Avestimehr_arXiv2015rank, Avestimehr_TIT14_TIM}. A low-rank
matrix completion approach with Riemannian algorithms has recently been proposed in \cite{Yuanming_2016LRMCTWC} to find the
minimum channel uses to achieve feasibility for {{any}} network topology.

In contrast, in this paper, we propose a different viewpoint: {{given any network topology and DoF allocation, we aim at finding the maximum number of admitted users to achieve the feasibility of topological interference management.}} We call this problem as \emph{user admission control} in topological interference management. User admission control is critical in wireless communication networks (i.e., cognitive radio access networks \cite{Tan_JSAC2014}, heterogeneous networks \cite{Tony_2015heterogeneous} and Cloud-RAN \cite{Yuanming_JSAC2015}) when quality-of-services (QoS) requirements are unsatisfied or the channel conditions are unfavorable \cite{Luo_2008useradmission}. Although the user admission control problems are normally non-convex mixed combinatorial optimization problems, a large body of recent work has demonstrated the effectiveness of convex relaxation for solving such problems \cite{Tan_JSAC2014, Tony_2015heterogeneous, Yuanming_JSAC2015, Luo_2008useradmission} based on the sum-of-infeasibilities in optimization theory \cite{boyd2004convex}. This is achieved by relaxing the original non-convex $\ell_0$-norm minimization problem for user admission control to the convex $\ell_1$-norm minimization problem \cite{boyd2004convex, Bach_ML2011}.

Unfortunately, the user admission control problem in topological interference management turns out to be highly intractable, which needs to optimize over continuous and combinatorial variables. To address the intractability, in this paper, we propose a sparse and low-rank modeling framework to compute the proposed solutions within polynomial time. In this model, sparsity  of the diagonal entries of the matrix (i.e.,
the number of non-zero entries) represents the number of the admitted users. The fixed low-rank constraint indicates the DoF allocation \cite{Yuanming_2016LRMCTWC}. However, the unique challenges arise in the proposed sparse and low-rank optimization model including the non-convex fixed-rank constraint and user capacity maximization objective function, i.e., $\ell_0$-norm objective \emph{maximization}. Novel algorithms thus need to be developed.

\subsection{Related Works}
\subsubsection{User Admission Control}
In dense wireless networks, user admission control is critical to maximize the user capacity while satisfying the QoS requirements for all the admitted users. To address the NP-hardness of the mixed combinatorial optimization problem, sparse optimization (e.g., $\ell_0$-norm minimization) approach, supported by the efficient algorithms (e.g., $\ell_1$-norm convex relaxation \cite{Tony_2015heterogeneous, Tan_JSAC2014} and the iterative reweighted $\ell_2$-algorithm \cite{Yuanming_JSAC2015}), provided an efficient way to find high quality solutions. However, convex
relaxation approach is inapplicable in our sparse and low-rank optimization problem due to the $\ell_0$-norm maximization as the objective. For the $\ell_1$-norm relaxation approach, it \changeBMM{yields} a $\ell_1$-norm maximization problem, which is still non-convex. Furthermore, maximizing $\ell_1$-norm shall yield unbounded values.     

\subsubsection{Low-Rank Models}
Low-rank models  \cite{Romberg_JSTSP16lrmc, Boyd_2014generalizedLowRank} inspire enormous applications  in machine learning, recommendation systems, sensor localization, etc. Due to the non-convexity of low-rank constraint or objective, many heuristic algorithms with optimality guarantees have been proposed in the last few years. In particular, convex relaxation approach using nuclear norm  \cite{Candes_2009exactMC} provides a polynomial time complexity algorithm with optimality guarantees via convex geometry and conic integral geometry analysis \cite{Tropp_livingedge2014}. 

The other popular way for low-rank optimization is based on matrix factorization, e.g., the alternating minimization \cite{Jain_2013lowAltmin, Boyd_2014generalizedLowRank} and Riemannian optimization method \cite{Vandereycken2013low}. In particular, the Riemannian optimization approach requires the smoothness of the objective function, while the alternating approach requires the convexity of the objective function. However, due to the non-convex and non-smooth objective function, we can not directly apply the existing matrix factorization approaches to solve the proposed sparse and low-rank optimization framework for user admission control.

Based on the above discussions, in contrast to the previous works on user admission control \cite{Tan_JSAC2014,
Tony_2015heterogeneous, Yuanming_JSAC2015, Luo_2008useradmission} and low-rank optimization problems \cite{Romberg_JSTSP16lrmc, Jain_2013lowAltmin, Boyd_2014generalizedLowRank, Vandereycken2013low},  we need to address the following coupled challenges to solve the sparse and low-rank optimization for user admission control in topological interference management:
\begin{itemize}
\item The objective of \emph{maximizing} the non-convex $\ell_0$-norm to maximize the user capacity, i.e., the number of admitted users;
\item Non-convex fixed-rank constraint to achieve a certain amount of DoF allocation.
\end{itemize}  

Therefore, unique challenges arise in the user admission control problem for topological interference management. We need to re-design the sparsity-inducing function and the efficient approach to deal with the fixed-rank constraint. 

\subsection{Contributions}
In this paper, we propose a sparse and low-rank optimization framework for user admission
control in topological interference management. The  Riemannian trust-region  algorithm is developed to solve the  proposed regularized smoothed $\ell_1$-norm sparsity inducing minimization problem, thereby guiding user selection. The main contributions are summarized as follows:
\begin{enumerate}
\item We propose a novel sparse and low-rank optimization framework to maximize the number of admitted users for achieving the feasibility of topological interference management. 

\item To avoid unboundness in the relaxed  $\ell_1$-norm maximization problem, a regularized  smoothed $\ell_1$-norm is proposed to induce sparsity pattern with bounded values, thereby guiding user selection.

\item  A  Riemannian trust-region algorithm is developed to solve the resulting rank-constrained smooth optimization problem for sparsity inducing. This is achieved by exploiting the quotient manifold of fixed-rank matrices. 
\item Simulation results \changeBMM{demonstrate} the effectiveness and near-optimal performance of the proposed Riemannian algorithm to maximize the user capacity for topological interference management. 

\end{enumerate}

\subsection{Organization}
The remainder of the paper is organized as follows. Section {\ref{sys}} presents the system model and problem formulation. A sparse and low-rank optimization framework for user admission control is proposed in Section {\ref{splrm}}. The Riemannian optimization algorithm is developed in Section \ref{roalg}. The ingredients of optimization on quotient manifold are presented in Section \ref{oqm}. Numerical results \changeBMM{are} illustrated in Section {\ref{simres}}. Finally, conclusions and discussions are presented in Section {\ref{condis}}.

\subsubsection*{Notations} Throughout this paper, $\|\cdot\|_p$ is the $\ell_p$-norm.
Boldface lower case and upper case letters represent vectors and matrices,
respectively. $(\cdot)^{-1}, (\cdot)^T, (\cdot)^{\sf{H}}$ and ${\rm{Tr}}(\cdot)$
denote the inverse, transpose, Hermitian and trace operators, respectively.
We use $\mathbb{C}$ and $\mathbb{R}$ to represent complex domain and real domain, respectively. $\mathbb{E}[\cdot]$ denotes
the expectation of a random variable. $|\cdot|$ stands for either the size
of a set or the absolute value of a scalar, depending on the context. We
denote ${\bm{A}}={\rm{diag}}\{x_1,\dots, x_N\}$ and ${\bm{I}}_N$ as a diagonal
matrix of order $N$ and the identity matrix of order $N$, respectively.

\section{System Model and Problem Formulation}
In this section, we present the channel model, followed by the user admission control problem to achieve the feasibility of topological interference management. 
\label{sys}
\subsection{Channel Model}
Consider the topological interference management problem in the partially connected $K$-user interference channel with each node quipped with a single antenna \cite{Jafar_TIT2013TIM,Yuanming_2016LRMCTWC}. Let $\mathcal{V}$ be the index set of the connected transceiver pairs such that the channel coefficient $h_{ij}$ between the transmitter $j$ and receiver $i$ is non-zero if $(i,j)\in\mathcal{V}$, and is zero otherwise. Each transmitter $i$ wishes to send a message
$W_i$ to its corresponding receiver $i$. The message $W_i$ is encoded into a vector $\bm{x}_i\in\mathbb{C}^r$ of length $r$. Therefore, over the $r$ channel uses, the received signal $\bm{y}_i\in\mathbb{C}^r$
at receiver $i$ is given by
\setlength\arraycolsep{2pt} 
\begin{eqnarray}
\bm{y}_i=h_{ii}\bm{x}_i+\sum_{i,j\in\mathcal{V}, i\ne j} h_{ij}\bm{x}_j+\bm{z}_i,
\forall i=1,\dots, K,
\end{eqnarray}
where $\bm{z}_i\sim\mathcal{CN}(\bm{0}, \bm{I}_r)$ is the additive noise
at receiver $i$. We consider the block fading channel, where the channel
coefficients stay constant during transmission, i.e., the channel coherence time is larger than channel uses $r$ for transmission. We assume each transmitter has an average power constraint, i.e., $\mathbb{E}\left[\|\bm{x}\|^2\right]\le
r \rho$ with $\rho>0$ as the maximum average transmit power.

The rate tuple $(R_1,\dots, R_K)$ is said to be achievable if there exists a  $(2^{rR_1},\dots, 2^{rR_K}, r)$ code scheme such that the average decoding error probability is vanishing as the code length $r$ approaches infinity. Here, we assume that each message $W_k$ is uniformly and independently chose over the $K$ message sets $\mathcal{W}_k:=\left[1:2^{rR_k}\right]$. In this paper, we choose  our performance metric as the symmetric DoF \cite{Jafar_TIT2013TIM,
Gesbert_TIT2015TIM}, i.e., the highest DoF achieved by all the users simultaneously,
\begin{eqnarray}
d_{\textrm{sym}}=\limsup_{\rho\rightarrow\infty}\sup_{(R_{\textrm{sym}},\dots, R_{\textrm{sym}})\in\mathcal{C}}{{R_{\textrm{sym}}}\over{\log \rho}},
\end{eqnarray}  
where $\mathcal{C}$ is the capacity region defined as the set of all the achievable rate tuples. The metric of DoF gives the first-order measurement of data rates \cite{tse2005fundamentals}.

\subsection{Topological Interference Management}   
In this paper, we restrict the class of the linear interference management strategies \cite{Jafar_IT2008, Jafar_TIT2013TIM, Yuanming_2016LRMCTWC}. Specifically, each transmitter $i$ encodes its message $W_i$ by a linear precoding vector $\bm{v}_i\in\mathbb{C}^r$ over $r$ channel uses:
\begin{eqnarray}
\bm{x}_i=\bm{v}_is_i,
\end{eqnarray} 
where $s_i\in\mathbb{C}$ is the transmitted data symbol. Here the precoding vectors $\bm{v}_i$'s only depend on the knowledge of network topology $\mathcal{V}$.  In this paper, we assume that the network connectivity information $\mathcal{V}$ is available at the transmitters. Therefore, over the $r$ channel uses, the received signal $\bm{y}_i\in\mathbb{C}^r$ at receiver $i$ can be rewritten as
\begin{eqnarray}
\bm{y}_i=h_{ii}\bm{v}_is_i+\sum_{i,j\in\mathcal{V}, i\ne j} h_{ij}\bm{v}_js_j+\bm{z}_i,
\forall i=1,\dots, K.
\end{eqnarray}

Let $\bm{u}_i\in\mathbb{C}^r$ be the decoding vector for each message
$W_i$ at receiver $i$. In the regime of asymptotically high signal-to-noise ratio (SNR), to accomplish decoding, we impose
the following interference alignment condition \cite{Jafar_IT2008, Jafar_TIT2013TIM, Yuanming_2016LRMCTWC} for
the precoding and decoding vectors:
\begin{eqnarray}
\label{c1}
{\bm{u}}_k^{\sf{H}}{\bm{v}}_k&\ne& 0, \forall k=1,\dots, K,\\
\label{c2}
{\bm{u}}_k^{\sf{H}}{\bm{v}}_i&=&0, \forall i\ne k, (i,k)\in\mathcal{V},
\end{eqnarray}
where the first condition is to preserve the desired signal and the second
condition is to align and cancel the interference signals. If conditions (\ref{c1})
and (\ref{c2}) are satisfied, the parallel interference-free channels can
be obtained over $r$ channel uses. Therefore, the symmetric DoF of $1/r$ is achieved for each message $W_i$ \cite{Jafar_TIT2013TIM}. We call this problem as \emph{topological interference management} \cite{Jafar_TIT2013TIM}, as only network topology information is required to establish the interference alignment conditions. 

However, establishing the conditions on $r$, $K$ and $\mathcal{V}$ to achieve feasibility of the interference alignment conditions (\ref{c1}) and (\ref{c2})
is challenging. In particular, given a number of users $K$ and channel uses $r$ (or DoF allocation $1/r$), the index coding approach \cite{Jafar_TIT2013TIM} and graph theory \cite{Gesbert_TIT2015TIM, Avestimehr_TIT14_TIM, Avestimehr_arXiv2015rank} were adopted to establish the conditions on the network topologies $\mathcal{V}$ to achieve feasibility for the interference alignment conditions  (\ref{c1}) and (\ref{c2}). The low-rank matrix completion approach \cite{Yuanming_2016LRMCTWC} has recently been proposed to find the minimum number of channel uses $r$ satisfying conditions (\ref{c1}) and (\ref{c2}), given any network topology information $\mathcal{V}$ and the number of uses $K$. The feasibility conditions of antenna configuration for interference alignment in MIMO interference channel has also been extensively investigated using algebraic geometry \cite{Jafar_TSP2010feasibility, Lau_TSP13IA, Tse_TIT2014feasibilityIA}.

In this paper, we put forth a different point of view on the feasibility conditions of topological interference management: given a number of $K$ users with any network topology $\mathcal{V}$ and the symmetric DoF allocation $1/r$, we \changeBMM{present} a novel {user admission control} approach to find the maximum number of the admitted users while satisfying the interference alignment conditions (\ref{c1}) and (\ref{c2}). Although user admission control has been extensively investigated in the scenarios of multiuser coordinated beamforming \cite{Luo_2008useradmission}, cognitive radio networks \cite{Tan_JSAC2014}, heterogeneous cellular networks \cite{Tony_2015heterogeneous} and Cloud-RAN \cite{Yuanming_JSAC2015}, this is the first time using the principle of user admission control in the framework of topological interference management. This shall provide a systematic framework for efficient algorithms design, as well as provide numerical insights into this challenging problem of topological interference management.

\section{A Sparse and Low-Rank Optimization Framework for  User Admission Control}
\label{splrm}
In this section, we present a user admission control approach to maximize the user capacity, i.e., find the maximum number of admitted users while satisfying the interference alignment conditions (\ref{c1})
and (\ref{c2}). This viewpoint is different from the previous works on finding the conditions of network topologies to achieve the feasibility of interference alignment \cite{Jafar_TIT2013TIM, Avestimehr_TIT14_TIM, Gesbert_TIT2015TIM, Avestimehr_arXiv2015rank}.

\subsection{Feasibility of Interference Alignment}
Given any network connectivity information $\mathcal{V}$ for the partially connected $K$-user interference channel, we say that the symmetric DoF allocation $1/r$ is feasible if there exists precoding vectors $\bm{v}_i\in\mathbb{C}^r$ and decoding vectors $\bm{u}_i\in\mathbb{C}^r$ such that the interference alignment conditions (\ref{c1}) and (\ref{c2}) are satisfied. Specifically, the feasibility of topological interference management problem can be formulated as 
\begin{eqnarray}
{\label{eq:problem_P_feasibility}}
\mathscr{F}: \mathop {\sf{find}}&&
\{{\bm{v}}_i\}, \{\bm{u}_i\}\nonumber\\
\subj&&{\bm{u}}_i^{\sf{H}}{\bm{v}}_i\ne 0, \forall i=1,\dots, K,\nonumber\\
&& {\bm{u}}_i^{\sf{H}}{\bm{v}}_j=0, \forall (i,j)\in\mathcal{V},
\end{eqnarray} 
where $\bm{v}_i\in\mathbb{C}^r$ and $\bm{u}_i\in\mathbb{C}^r$ are optimization variables.

However, the solutions to the feasibility problem (\ref{eq:problem_P_feasibility}) is unknown in general. In particular, the index coding approach \cite{Jafar_TIT2013TIM} and the graph theory \cite{Gesbert_TIT2015TIM, Avestimehr_TIT14_TIM, Avestimehr_arXiv2015rank} were adopted to establish the conditions on the network topology $\mathcal{V}$ to achieve feasibility of interference alignment. On the other hand, the low-rank matrix completion approach was proposed in \cite{Yuanming_2016LRMCTWC} to find the minimum number of channel uses $r$ to achieve interference alignment feasibility for any  network topology $\mathcal{V}$.

In contrast, in this paper, our goal is to maximize the user capacity, i.e., the find the maximum number of admitted users while satisfying the interference alignment conditions:
\begin{eqnarray}
{\label{eq:problem_P11}}
\mathop {\maxi}\limits_{\{\bm{v}_i\}, \{\bm{u}_i\}}&&
|\mathcal{S}|\nonumber\\
\subj&&{\bm{u}}_i^{\sf{H}}{\bm{v}}_i\ne 0, \forall i\in\mathcal{S}, \nonumber\\
&& {\bm{u}}_i^{\sf{H}}{\bm{v}}_j=0, \forall i\ne j, i,j\in\mathcal{S}, (i,j)\in\mathcal{V},
\end{eqnarray}
where $\mathcal{S}\subseteq\{1,\dots, K\}$ is the admitted users, $\bm{v}_i\in\mathbb{C}^r$ and $\bm{u}_i\in\mathbb{C}^r$. This problem is called as the \emph{user admission control problem}. Unfortunately, it turns out to be highly intractable due to the non-convex quadratic constraints and the non-convex combinatorial objective function. To assist efficient algorithms design, in this paper, we propose a sparse and low-rank optimization for user admission control via exploiting the sparse and low-rank structures in problem (\ref{eq:problem_P11}).

\subsection{Sparse and Low-Rank Optimization Paradigms for User Admission Control}
\begin{figure}[t]
\center
\includegraphics[scale = 0.4]{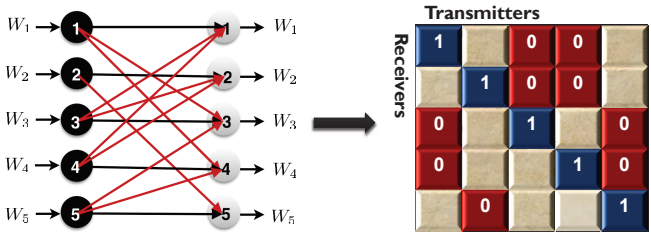}
\caption{(a) The topological interference alignment problem for the partially
connected $K$-user interference channel with  only the knowledge of the network
connectivity information available. The interference links are marked as
red while the desired links are marked as black. (b) The corresponding incomplete
matrix with ``0" indicating interference alignment and cancellation and ``1"
representing desired signal preserving.}
\label{LRMC}
\end{figure}

Let $\bm{X}=[X_{ij}]\in\mathbb{C}^{K\times K}$ with
$X_{ij}={\bm{u}}_i^{\sf{H}}{\bm{v}}_j\in\mathbb{C}$. 
The interference alignment conditions (\ref{c1}) and (\ref{c2}) thus can be rewritten as
\begin{eqnarray}
%\label{C1}
X_{kk}&\ne& 0, \forall k=1,\dots, K,\\
%\label{C2}
X_{ki}&=&0, \forall i\ne k, (i,k)\in\mathcal{V}.
\end{eqnarray}
For other entries $X_{ki}, \forall (k,i)\notin\mathcal{V}$, they can be any values. Observing that the achievable symmetric DoF is given
by
\begin{eqnarray}
{\sf{DoF}}=1/{\rm{rank}}(\bm{X})=1/r,
\end{eqnarray}
a low-rank matrix completion problem was proposed in \cite{Yuanming_2016LRMCTWC} to find the minimum channel uses while satisfying the interference alignment conditions.
Fig. {\ref{LRMC}} demonstrates the procedure of transforming the topological interference alignment conditions (\ref{c1}) and (\ref{c2}) into the associated incomplete matrix $\bm{X}$.

Define $\bm{X}(\mathcal{S})\in\mathbb{C}^{|\mathcal{S}|\times|\mathcal{S}|}$
as the submatrix of
$\bm{X}$, i.e., $\bm{X}(\mathcal{S})=[X_{ij}]_{i,j\in\mathcal{S}}$. The rank of the submatrix $\bm{X}(\mathcal{S})$ equals $r$. The user admission control problem (\ref{eq:problem_P11}) can be further reformulated as follows: 
\begin{eqnarray}
{\label{eq:problem_P}}
\mathop {\maxi}\limits_{{\bm X} \in \mathbb{C}^{K\times K},
\mathcal{S}}&&
|\mathcal{S}|\nonumber\\
\subj&&{\rm{rank}}({\bm{X}}(\mathcal{S}))=r, \nonumber\\
&& X_{ii}\ne 0, \forall i\in\mathcal{S},\nonumber\\
&& X_{ij}=0, \forall i\ne j, i,j\in\mathcal{S}, (i,j)\in\mathcal{V},
\end{eqnarray}
where the first constraint is to preserve the  symmetric DoF allocation as $1/r$. However,
problem (\ref{eq:problem_P}) is still a highly intractable mixed
combinatorial optimization
problem with a non-convex fixed-rank constraint and a combinatorial objective function. 

To enable the capability of polynomial-time complexity algorithm design, we further reveal the sparsity structure in problem (\ref{eq:problem_P}) for user admission control. We notice that 
\begin{eqnarray}
\label{zerouse}
\|{\diagg}(\bm{X})\|_0=|\mathcal{S}|,
\end{eqnarray}
where $\diagg(\cdot)$ extracts the diagonal of a matrix and $\|\cdot \|_0$
is the $\ell_0$-norm of a vector, i.e., the count of non-zero entries. Problem (\ref{eq:problem_P}) can be further reformulated
as the following sparse and low-rank optimization
problem, i.e.,
\begin{eqnarray}\label{eq:ell0}
\mathscr{P}:\mathop {\maxi}\limits_{{\bm X} \in \mathbb{R}^{K\times K}}&&
\|{\diagg}(\bm{X})\|_0\nonumber\\
\subj&&{\rm{rank}}({\bm{X}})=r,\nonumber\\
\label{affcon}
&& X_{ij}=0, \forall i\ne j, (i,j)\in\mathcal{V}.
\end{eqnarray}
Notice that we only need to consider problem $\mathscr{P}$ in the real field without losing any performance in terms of admitted users. The reason is that the affine constrain (\ref{affcon}) is restricted in real field and the diagonal entries of matrix $\bm{X}$ can be further restricted to the real field while achieving the same value of $\|{\diagg}(\bm{X})\|_0$ in the complex field.

Sparse optimization has shown to be powerful for the user admission problems \cite{Luo_2008useradmission, Tan_JSAC2014,
Tony_2015heterogeneous, Yuanming_JSAC2015} via $\ell_0$-norm \emph{minimization} using the sum-of-infeasibilities convex relaxation heuristic in optimization theory \cite[Section 11.4]{boyd2004convex}. In particular, to maximize the number of admitted users is equivalent to minimize the number of  violated inequalities for the quality-of-service (QoS) constraints. Although problem $\mathscr{P}$ adopts the same  philosophy of $\ell_0$-norm to count the number of admitted users (\ref{zerouse}), it reveals unique challenges due to  $\ell_0$-norm \emph{maximization} and non-convex fixed-rank constraint. However, compared with the original formulation (\ref{eq:problem_P}), the  sparse and low-rank optimization formulation (\ref{eq:ell0}) holds algorithmic advantages, which \changeBMM{are} demonstrated in the sequel via the Riemannian optimization approach \cite{Absil_2009optimizationonManifolds}.

\subsection{Problem Analysis}
In this subsection, we reveal the unique challenges of solving the sparse and low-rank optimization problem $\mathscr{P}$ for user admission control in topological interference management.    
\subsubsection{Non-convex Objective Function}
\begin{figure}[t]
\center
\includegraphics[scale = 0.45]{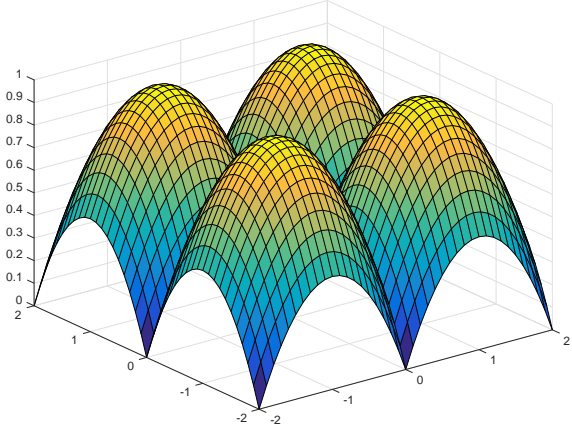}
\caption{The regularized sparsity inducing norm $f(\bm{z})=\|\bm{z}\|_1-0.5\|\bm{z}\|_2^2$
with bounded values in $\bm{z}\in\mathbb{R}^2$.}
\label{3d}
\end{figure}
Although $\ell_1$-norm serves the convex surrogate for the non-convex $\ell_0$-norm \cite{boyd2004convex, Bach_ML2011}, it is inapplicable in problem $\mathscr{P}$ for $\ell_0$-norm \emph{maximization}, as it yields unbounded values. To aid efficient algorithms design, we \changeBMM{propose} a novel regularized $\ell_1$-norm to induce sparsity with bounded values. This is achieved by adding a quadratic term in the $\ell_1$-norm as follows:
\begin{eqnarray}
\label{rsnorm}
f(\bm{z})=\|\bm{z}\|_1-\lambda\|\bm{z}\|_2^2,
\end{eqnarray}
where $\bm{z}\in\mathbb{R}^n$ and  $\lambda \ge0$ is a weighting parameter. A typical example with $f(\bm{z})=\|\bm{z}\|_1-0.5\|\bm{z}\|_2^2$ and $\bm{z}\in\mathbb{R}^2$ is illustrated in Fig. {\ref{3d}}, which upper
bounds all the diagonal values by $1$.

\subsubsection{Non-convex Fixed-rank Constraint}
Matrix factorization serves a powerful way to address the non-convexity of the fixed-rank matrices. One popular way is to factorize a fixed rank-$r$ matrix $\bm{X}$ (\changeBMM{in real field}) as $\bm{U}\bm{V}^{{T}}$ with $\bm{U}\in\mathbb{R}^{K\times r}$ and $\bm{V}\in\mathbb{R}^{K\times r}$, followed by alternatively optimizing over $\bm{U}$ and $\bm{V}$ holding the other fixed \cite{Boyd_2014generalizedLowRank, Jain_2013lowAltmin}. However, due to the non-convex objective function in problem $\mathscr{P}$, the resulting optimization problem over $\bm{U}$ or $\bm{V}$ is still non-convex. Furthermore, such factorization is not unique as $\bm{X}$ remains unchanged under the transformation of the factors
\begin{eqnarray}
(\mat{U},\mat{V})\mapsto (\mat{U}\mat{M}^{-1},\mat{V}\mat{M}^{T}),
\end{eqnarray}
for all non-singular matrices $\mat{M} $ of size $r\times r$. As a result, the critical points of an objective function \changeBM{parameterized
with $\mat U$ and $\mat V$} are \emph{not isolated} on $\mathbb{R} ^{K \times r} \times \mathbb{R} ^{K \times r}$. \changeBMM{This profoundly affects the performance of second-order optimization algorithms which require non degenerate critical points, which is no longer the case here. We propose to address this issue by exploiting the \emph{quotient manifold geometry} of the set of fixed-rank matrices \cite{Mishra_2014fixedrank}. The resulting non-convex optimization problem is further solved by exploiting the Riemannian optimization framework which provides systematic ways to develop algorithms on quotient manifolds \cite{Absil_2009optimizationonManifolds}.}  

In summary, in this paper, we propose a new powerful approach to induce the sparsity in the solution to problem $\mathscr{P}$, followed by the Riemannian optimization approach via exploiting the quotient manifold geometry of fixed-rank matrices. The induced sparsity pattern guides user selection for user admission control.

\section{Regularized Smoothed $\ell_1$-Minimization for Sparse and Low-Rank Optimization via Riemannian Optimization}
\label{roalg}
\begin{figure}[t]
\center
\includegraphics[scale = 0.45]{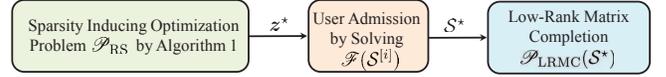}
\caption{The proposed three-stage Riemannian framework for user admission control in topological interference alignment via sparse and low-rank optimization. $\bm{z}^{\star}\in\mathbb{R}^K$ is the induced sparsity pattern for user selection and $\mathcal{S}^{\star}\subseteq\{1,\dots, K\}$ is set of admitted users.}
\label{slopt}
\end{figure}

In this section, we present a Riemannian framework for sparse and low-rank optimization problem $\mathscr{P}$ via regularized smoothed $\ell_1$-minimization by exploiting the quotient manifold geometry of fixed-rank matrices. The induced sparsity solution to problem $\mathscr{P}$ provides guideline for user admission control, supported by a user selection procedure. In the final stage, a low-rank matrix completion approach with Riemannian optimization \changeBMM{is} adopted to design the linear topological interference management strategy. The proposed three-stage Riemannian framework for user admission control in topological interference management is presented in Fig. {\ref{slopt}}.    

\subsection{Stage One: Regularized Smoothed $\ell_1$-Minimization for Sparsity Inducing}
In order to make problem $\mathscr{P}$ (\ref{eq:ell0}) numerically tractable, we relax
the non-convex $\ell_0$-norm objective function to its convex surrogate $\ell_1$-norm, resulting in the following optimization problem:
\begin{equation}\label{eq:ell1}
\begin{array}{lll}
\mathop {\maxi}\limits_{{\bm X} \in \mathbb{R}^{K\times K}}&&
\|{\diagg}(\bm{X})\|_1 \\
\subj&&{\rankk}({\bm{X}})=r, \\
&& {X}_{ij}=0, \forall i\ne j, (i,j)\in\mathcal{V}.
\end{array}
\end{equation}
Although the $\ell_1$-norm is tractable, it is \emph{unbounded} from above due to $\ell_1$-norm \emph{maximization},
which makes problem (\ref{eq:ell1}) ill-posed. Note that maximizing a convex $\ell_1$-norm is still non-convex. 

To circumvent the unboundness issue,
we add the quadratic term $-  \lambda \|{\diagg}(\bm{X})\|_2^2$ to the objective function
in problem (\ref{eq:ell1}), where $\lambda \ge0$ is a weighting parameter that bounds
the overall objective function from above leading to the formulation
\begin{equation}\label{eq:ell1_mod}
\begin{array}{lll}
\mathop {\maxi}\limits_{{\bm X} \in \mathbb{R}^{K\times K}}&&
\|{\diagg}(\bm{X})\|_1 - \lambda \|{\diagg}(\bm{X})\|_2^2  \\
\subj&&{\rankk}({\bm{X}})=r, \\
&& {X}_{ij}=0, \forall i\ne j, (i,j)\in\mathcal{V}.
\end{array}
\end{equation}
For example, if $\lambda = 0.5$, then the diagonal values of $\mat{X}$ are
upper bounded by $1$. It should be emphasized that the role of $\lambda$
in (\ref{eq:ell1_mod}) is to upper bound the objective function and it does
not affect the sparsity pattern that is expected from (\ref{eq:ell1}). This \changeBMM{is} further be confirmed in Section {\ref{sec:fixed_rank}} via simulations. \changeBMM{Additionally,
if $\mat{X}^\star$ is the solution to (\ref{eq:ell0}), then $\alpha \mat{X}^\star$
is also a solution of (\ref{eq:ell0}) for all non-zero scalar $\alpha$. Equivalently,
there exists continuum of solutions, which is effectively resolved by the
objective function in (\ref{eq:ell1_mod})}. 

Although problem (\ref{eq:ell1_mod}) is still non-convex due to the non-convex objective (i.e., maximizing a convex function) and non-convex fixed-rank constraint, it has the algorithmic advantage \changeBMM{that it} can be solved efficiently \changeBMM{(i.e., numerically)} in the framework of Riemannian optimization \cite{Absil_2009optimizationonManifolds}.

\subsubsection{Riemannian Optimization for Fixed-Rank Optimization}
In this subsection, we propose a Riemannian optimization algorithm to solve the non-convex optimization problem (\ref{eq:ell1_mod}), which is equivalent to  
\begin{equation}\label{eq:ell1_mod_minimize}
\begin{array}{lll}
\mathop {\mini}\limits_{{\bm X} \in \mathbb{R}^{K\times K}}&&
-\|{\diagg}(\bm{X})\|_1 + \lambda \|{\diagg}(\bm{X})\|_2^2  \\
\subj&&{\rankk}({\bm{X}})=r, \\
&& {X}_{ij}=0, \forall i\ne j, (i,j)\in\mathcal{V}.
\end{array}
\end{equation}
\changeBM{However, the intersection of rank constraint and the affine constraint is challenging to characterize}. We, \changeBM{therefore}, propose to solve problem (\ref{eq:ell1_mod_minimize}) via a \emph{regularized} version as follows:
\begin{eqnarray}\label{eq:regularized_formulation}
\mathscr{P}_{\textrm{RS}}:\mathop {\mini}\limits_{{\bm X} \in \mathbb{C}^{K\times K}}&&
\displaystyle\frac{1}{2}\!\!\!\!\!\!\underbrace{\sum\limits_{(i,j) \in \mathcal{V}}
\!\!\!{X}_{ij}
^2}_{\rm network\ topology } \!\!\!\!\!\!\!+ \rho \underbrace{ \sum\limits_{i=1}^K(\lambda
 X_{ii}^2 - ({X}_{ii}^2 + \epsilon^2)^{1/2})}_{\rm admission}\nonumber\\
\subj&&{\rm{rank}}({\bm{X}})=r,
\end{eqnarray}
where $\rho\ge 0$ is the regularization parameter and $\epsilon$ is the parameter
that approximates $|{X}_{ii}|$ with the smooth term $\left({X}_{ii}^2 + \epsilon
^2\right)^{1/2}$ that makes the objective function \emph{differentiable}.
A very small $\epsilon$ leads to ill-conditioning of the objective function
in (\ref{eq:regularized_formulation}). Since we intend to obtain the sparsity pattern of the optimal $\bm X$, we set $\epsilon$ to a high value, \changeBM{e.g., $0.01$}, to make problem (\ref{eq:regularized_formulation}) well conditioned. Problem $\mathscr{P}_{\textrm{RS}}$ is an optimization problem over the set of fixed-rank matrices and can be solved
via a Riemannian trust-region algorithm \cite{Absil_2009optimizationonManifolds}.

%Similarly, a larger $\rho$ induces more sparsity in $\bm{X}$. 

\subsection{Stage Two: Finding Sparsity Pattern for User Admission Control}

%In practice, a good choice of $\rho$ is $0.001$, which is obtained by cross-validation. 

%It should be noted that the problem (\ref{eq:regularized_formulation}) is minimizing a smooth objective function over the set of fixed-rank matrices for which numerous algorithms exist including second-order optimization algorithms.

Let $\bm{X}^{\star}$ be the solution to the regularized smoothed $\ell_1$-minimization problem $\mathscr{P}_{\textrm{RS}}$. We order the diagonal entries of matrix $\bm{X}^{\star}$, i.e., the vector ${\bm{z}}^{\star}={\rm{diag}}(\bm{X}^{\star})\in\mathbb{R}^K$,  in the descending order: $|z_{\pi_1}|\ge |z_{\pi_2}|\ge\cdots\ge |z_{\pi_{K}}|$. The user with larger coefficients $z_i$ \changeBMM{has} a higher priority to be admitted. We adopt the bi-section search procedure to find the maximum number of admitted users. Specifically, let $N_0$ be the maximum number of users that can be admitted while satisfying the interference alignment conditions. To determine the value of $N_0$, a sequence of the following size-reduced topological interference management feasibility problem needs to be solved,
\begin{eqnarray}
\label{uafeasible}
\mathscr{F}(\mathcal{S}^{[m]}):\mathop {\sf{find}}&&
\bm{X}(\mathcal{S}^{[m]})\nonumber\\
\subj&&{\rm{rank}}({\bm{X}}(\mathcal{S}^{[m]}))=r,\nonumber\\
&& X_{ii}=1, \forall i\in\mathcal{S}^{[m]},\\
&& X_{ij}=0, \forall i\ne j, i,j\in\mathcal{S}^{[m]}, (i,j)\in\mathcal{V},\nonumber
\end{eqnarray}
where $\mathcal{S}^{[m]}=\{\pi_1,\dots, \pi_m\}$.

To check the feasibility, we rewrite problem (\ref{uafeasible}) as follows:
\begin{eqnarray}
\label{ls}
\mathop {\mini}&&
\|\mathcal{P}_{\Omega}(\bm{X}(\mathcal{S}^{[m]}))-\bm{I}_{|\mathcal{S}^{[m]}|}\|_F^2\nonumber\\
\subj&&{\rm{rank}}({\bm{X}}(\mathcal{S}^{[m]}))=r,
\end{eqnarray}
where $\Omega=\{(i,j)|i,j\in\mathcal{S}^{[m]}, (i,j)\in\mathcal{V}\}$ and $\mathcal{P}_{\Omega}(\bm{Y}):\mathbb{R}^{n\times
n}\rightarrow\mathbb{R}^{n\times n}$ is
the orthogonal projection operator onto the subspace of matrices which vanish
outside $\Omega$ such that the $(i,j)$-th component of $\mathcal{P}_{\Omega}(\bm{Y})$
equals to $Y_{ij}$ if $(i,j)\in\Omega$ and zero otherwise. If the objective
value approaches to zero, we say that the set of users $\mathcal{S}^{[m]}$ can be admitted.
Problem (\ref{ls}) can be solved by Riemannian trust-region algorithms \cite{Mishra_RP2014}
via {Manopt} \cite{manopt}. \changeBMM{Note that, theoretically, the Riemannian algorithm can only guarantee convergence to a first-order critical point, but empirically, we observe convergence to critical points that are local minima.}

\subsection{Stage Three: Low-Rank Matrix Completion for Topological Interference Management}
Let $\mathcal{S}^{\star}=\{\pi_1,\dots, \pi_{N_0}\}$ be the admitted users. We need to solve the following sized-reduced rank-constrained \emph{matrix completion} problem:
\begin{eqnarray}
\label{eq:refining}
\mathscr{P}_{\textrm{LRMC}}(\mathcal{S}^{\star}):\mathop {\mini}&&
\|\mathcal{P}_{\Omega}(\bm{X}(\mathcal{S}^{\star}))-\bm{I}_{|\mathcal{S}^{\star}|}\|_F^2\nonumber\\
\subj&&{\rm{rank}}({\bm{X}}(\mathcal{S}^{\star}))=r,
\end{eqnarray} 
to find the precoding
vectors $\bm{v}_i$'s and decoding vectors $\bm{u}_i$'s for the admitted users in $\mathcal{S}^{\star}$.

Therefore, the proposed three-stage Riemannian optimization based user admission control algorithm is presented in Algorithm {\ref{algua}}.

\begin{algorithm}
\label{algua}
\caption{User Admission Control for Topological Interference Management via Riemannian Optimization}
\textbf{Step 0:} Solve the sparse inducing optimization
problem $\mathscr{P}_{\textrm{RS}}$ (\ref{eq:regularized_formulation}) using the Riemannian trust-region algorithm in Section {\ref{oqm}}.
Obtain the solution ${\bm{X}}^{\star}$ and  sort the diagonal entries in the descending
order: $|z_{\pi_1}|\ge\dots\ge |z_{\pi_{K}}|$, {\bf{go to Step 1}}.\\
\textbf{Step 1:} Initialize $N_{\textrm{low}}=0$, $N_{\textrm{up}}=K$, $i=0$.\\
\textbf{Step 2:} Repeat
\begin{enumerate}
\item Set $i\leftarrow\left\lfloor{{N_{\textrm{low}}+N_{\textrm{up}}}\over{2}}\right\rfloor$.\\
\item Solve  problem $\mathscr{F}(\mathcal{S}^{[i]})$ (\ref{uafeasible}) via (\ref{ls}) using the Riemannian trust-region algorithm in Section {\ref{oqm}}:
if \\it is feasible, set $N_{\textrm{low}}=i$; otherwise, set $N_{\textrm{up}}=i$.
\end{enumerate}
\textbf{Step 3:} Until $N_{\textrm{up}}-N_{\textrm{low}}=1$, obtain $N_{0}=N_{\textrm{up}}$
and obtain the  admitted users set $\mathcal{S}^{\star}=\{\pi_{1},\dots,
\pi_{N_0}\}$.\\
\textbf{Step 4:} Solve problem $\mathscr{P}_{\textrm{LRMC}}(
\mathcal{S}^{\star})$
(\ref{eq:refining}) to obtain the precoding and decoding vectors for the
admitted users.\\
\textbf{End}
\end{algorithm}

% \begin{table}[t]
% \caption{Riemannian Optimization Algorithm for $\mathscr{P}$.}
% \label{tab:overall_algorithm} 
% \begin{center}  \small
% \begin{tabular}{ |p{8.2cm}| }
% \hline
% 
% 
% %Riemannian optimization Algorithm for $\mathscr{P}$.
% 
% \begin{itemize}
% \setlength\itemsep{0.3em}
% \item Finding sparsity partition: we solve the \emph{regularized} formulation (\ref{eq:regularized_formulation}) to identify a good \emph{sparsity} pattern $\bm P$, which is a binary vector of size $K \times 1$ with $1$s at \changeBM{non-zero} positions and $0$s at \changeBM{zero positions}.
% 
% \item Refining: once the sparsity pattern $\bm P$ is determined, we solve the matrix completion problem (\ref{eq:refining}) with rank constraint to refine the estimate obtained from the regularized formulation solution. 
% 
% \item Both (\ref{eq:regularized_formulation}) and (\ref{eq:refining}) are solved with a Riemannian trust-region algorithm on the set of fixed-rank matrices. 
% \end{itemize}
% \\
% \hline
% \end{tabular}
% \end{center} 
% \end{table}  
% 
% 

\begin{figure}[t]
\center
\includegraphics[scale = 0.35]{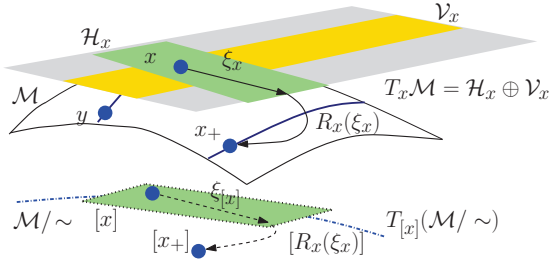}
\caption{Optimization on a quotient manifold. {The dotted lines represent abstract objects and the solid lines are their matrix representations.} The points $ {x}$ and $ y$ in the total (computational) space $ { \mathcal{M}}$ belong to the same equivalence class (shown in solid blue color) and they represent a single point $ [ {x}] :=  \{ {y} \in  {\mathcal M} :  {y} \sim  {x }\}$ in the quotient space $\mathcal{M}/\sim$. An algorithm by necessity is implemented in the computation space, but conceptually, the search is on the quotient manifold. Given a search direction $\xi_x$ at $x$, the updated point on $\mathcal{M}$ is given by the retraction mapping $R_x$.}
\label{fig:manifold_optimization}
\end{figure}

\subsection{The Framework of Fixed-Rank Riemannian Manifold Optimization}\label{sec:fixed_rank}
The optimization problems (\ref{eq:regularized_formulation}), (\ref{ls}), and (\ref{eq:refining}) are \emph{least-square} optimization problems with fixed rank constraint. A rank-$r$ matrix $\mat{X} \in \mathbb{R}^{K \times K}$ is parameterized as $\mat{X} = \mat{U} \mat{V}^T$, where $\mat{U} \in \mathbb{R} ^{K \times r}$ and $\mat{V} \in \mathbb{R}^{K
\times r}$ are full column-rank matrices. Such a factorization, however, is not unique as $\mat{X}$ remains unchanged under the transformation of the factors
\begin{equation}\label{eq:symmetry_gh}
(\mat{U},\mat{V})\mapsto (\mat{U}\mat{M}^{-1},\mat{V}\mat{M}^{T}),
\end{equation}
for all non-singular matrices $\mat{M} \in \GL{r}$, the set of $r \times r$ non-singular matrices. \changeBM{Equivalently, $\mat{X} = \mat{U} \mat{V}^T = \mat{U} \mat{M}^{-1}  (\mat{V}   \mat{M}^T)^T$ for all non-singular matrices $\mat M$.} As a result, the critical points of an objective function \changeBM{parameterized with $\mat U$ and $\mat V$} are \emph{not isolated} on $\mathbb{R} ^{K \times r} \times \mathbb{R} ^{K \times r}$.

The classical remedy to remove this indeterminacy requires further (triangular-like) structure in the factors $\mat{U}$ and $\mat{V}$. For example, LU decomposition is a way forward. In contrast, we encode the invariance map (\ref{eq:symmetry_gh}) in an abstract search space by optimizing directly over a set of equivalence classes 
\begin{equation}\label{eq:equivalence-classes-balanced}
 [(\mat{U},\mat{V})] := \{ (\mat{U}\mat{M}^{-1},\mat{V}\mat{M}^{T}): \mat{M}
\in\mathrm{GL}(r) \}.
\end{equation}
The set of equivalence classes is termed as the \emph{quotient space } and is denoted by 
\begin{equation}\label{eq:quotient-balanced}
        {\mathcal{M}}_r:= \mathcal{M} /\GL{r},
\end{equation}
where the total space ${\mathcal{M}}$ is the product space $\mathbb{R}^{K
\times r} \times \mathbb{R}^{K \times r}$.

Consequently, if an element $x \in \mathcal{M}$ has the matrix characterization $(\mat{U},
\mat{V})$, then (\ref{eq:regularized_formulation}), (\ref{ls}), and (\ref{eq:refining})
are of the form 
\begin{equation}\label{eq:manifold_formulation}
\begin{array}{lll}
\mathop {\sf minimize}\limits_{[x] \in \mathcal{M}_r} && f([x]),\\
\end{array}
\end{equation}
where $[x]=[(\mat{U},\mat{V})] $ is defined in (\ref{eq:equivalence-classes-balanced}) and $f: \mathcal{M} \rightarrow \mathbb{R}:x \mapsto f(x)$ is a \emph{smooth} function on $\mathcal{M}$, but now induced (with slight abuse of notation) on the quotient space $\mathcal{M}_r$ (\ref{eq:quotient-balanced}).

The quotient space $\mathcal{M}_r$ has the structure of a smooth \emph{Riemannian} quotient manifold of $\mathcal{M}$ by $\GL{r}$ \cite{Mishra_2014fixedrank}. The Riemannian structure conceptually transforms a rank-constrained optimization problem into an \emph{unconstrained} optimization problem over the non-linear manifold $\mathcal{M}_r$. Additionally, it allows to compute objects like gradient (of an objective function) and develop a \changeBM{Riemannian} trust-region algorithm on $\mathcal{M}_r$ that uses second-order information for faster convergence \cite{Absil_2009optimizationonManifolds}. 

%It should be noted that $\GL{r}$ takes away
%all the symmetry of the total space as the dimension of $\mathcal{M}_r$ is
%$(K+K-r)r$ which is equal to the dimension of the set of rank-$r$ matrices
%of size $K\times K$.

\begin{table*}[!htbp]
\caption{Manifold-related ingredients}
\label{tab:spaces} 
\begin{center} \small
\begin{tabular}{ p{7cm} | p{10cm} } 
& $\mat{X} = \mat{U}\mat{V}^T$  \\
\hline
&  \\

Matrix representation &  ${x} = (\mat{U}, \mat{V})$
   \\ 
&  \\

Total space ${\mathcal M}$ &  $\mathbb{R}^{K \times r} \times \mathbb{R}^{K \times r}$
  \\ 
 &   \\

Group action &  $(\mat{U}\mat{M}^{-1}, \mat{V}\mat{M}^T), \mat{M} \in  {\rm GL}(r)$
   \\
   & \\ 

Quotient space ${\mathcal M}/\sim$ &  
$
\begin{array}[t]{lll}
\mathbb{R}^{K \times r} \times \mathbb{R}^{K \times r} 
/  {\rm GL}(r) 
\end{array}
$
   \\

&  \\
Vectors in the ambient space
&  
$
(\mat{Z}_{\mat{U}}, \mat{Z}_{\mat{V}} )  \in  
\mathbb{R}^{K \times r} \times  \mathbb{R}^{K \times r}
$
\\
&   \\
Matrix representation of a tangent vector $\xi_x$ in $T_{x}{\mathcal M}$
&
$
({\xi}_{\mat{U}}, {\xi}_{\mat{V}} )  \in  
\mathbb{R}^{K \times r} \times  \mathbb{R}^{K \times r} 
$
   \\ 
&   \\
Metric 
$
{g}_{ x} ({\xi}_{x},  {\eta}_{x})$ for any $ {\xi}_{x}, {\eta}_{ x} \in T_{x} {\mathcal M}$
&
 $
 \begin{array}[t]{lll}
  \trace ( (\mat{V}^T\mat{V}) {\xi}^T_{\mat U} {\eta}_{\mat U}  )  
 +  \trace ( (\mat{U}^T\mat{U}) {\xi}^T_{\mat V}  {\eta}_{\mat V}  )
 \end{array}
 $
 \\
&  \\

Vertical tangent vectors in
$
\mathcal{V}_{ x}
$
& 
$
\begin{array}[t]{lll}
 \{  (   -\mat{U}\mat{\Lambda} , \mat{V}\mat{\Lambda} ^T   )  : 
  \mat{\Lambda} \in \mathbb{R}^{r\times r} \}
\end{array}
$
\\

&   \\
Horizontal tangent vectors in 
$
\mathcal{H}_{ x}
$
& 
$
  \{ ( { \zeta}_{\mat{U}} , {\zeta}_{\mat{V}}   )
  \in \mathbb{R}^{K \times r}  \times \mathbb{R}^{K \times r} :
    \mat{U}^T \zeta_{\mat{U}} \mat{V}^T \mat{V}  = \mat{U}^T\mat{U} \zeta_{\mat{V}}^T \mat{V}  \}
$
\\

&  \\

Projection of a tangent vector $\eta_x \in T_x \mathcal{M}$ on the horizontal space $\mathcal{H}_x$&  $\Pi_{x}({\eta}_{{x}})=  
( {\eta}_{\mat{U}} + \mat{U}\mat{\Lambda} , {\eta}_{\mat{V}} - \mat{V\Lambda}^T)$, where $\mat{\Lambda}  =   0.5 ( \eta_{\mat{V}}^T\mat{V} (\mat{V}^
T \mat{V})^{-1}  - (\mat{U}^T\mat{U} )^{-1}\mat{U}^T\eta_{\mat{U}} )$. \\

&  \\

Retraction of a horizontal vector ${\xi}_{x}$ onto the manifold &  
 
 $
\begin{array}[t]{lll}R_{ x} ({\xi}_{x}) = (\mat{U} +  {\xi}_{\mat U},  \mat{V} + {\xi}_{\mat V} )
 \end{array}
$ \\

&  \\

Matrix representation of the Riemannian gradient $\grad_{ x} f$
&
$(\frac{\partial f}{\partial
\mat{U}} (\mat{V}^T\mat{V})^{-1},  \frac{\partial f}{\partial
\mat{V}} (\mat{U}^T\mat{U})^{-1})$, where ${\partial f}/{\partial \mat{U}}$ and ${\partial f}/{\partial \mat{V}}$ are the {partial derivatives} of $f$ with respect to $\mat{U}$ and $\mat{V}$, respectively.
\\

& \\

Matrix representation of the Riemannian Hessian ${\hess_x} f [\xi_x]$ along a horizontal vector $\xi_x $  
&
$ \Pi_{{x}}(  {\rc}_{{\xi}_{x}} { {\grad_{ x}} f}   )$, where ${\grad_x} f$ has the representation shown above. The matrix representation of the Riemannian connection $\rc _{\xi_x} \eta_x$ is shown in (\ref{eq:connection_total_space}). Finally, the projection operator $\Pi_x$ is defined in (\ref{eq:projection_gh}).
\\
&   \\

\hline
\end{tabular}
\end{center} 
\end{table*}

\section{Optimization on Quotient Manifold}
\label{oqm}

Consider an equivalence relation $\sim$ in the \emph{total} (computational)
space $ {\mathcal{M}}$. The quotient manifold ${\mathcal M}/\sim$ generated by this equivalence property consists of elements that are \emph{equivalence classes} of the form $ [ {x}] =  \{ {y} \in  {\mathcal M} :  {y} \sim  {x}\}$. Equivalently, if $[x]$ is an element in $\mathcal{M}/\sim$, then its matrix representation in $\mathcal{M}$ is $x$. \changeBMM{In the context of rank constraint, ${\mathcal M}/\sim$ is identified with $\mathcal{M}_r$, i.e., the fixed-rank manifold}. Fig. \ref{fig:manifold_optimization} shows a schematic viewpoint of optimization on a quotient manifold. Particularly, we need the notion of ``linearization'' of the search space, ``search'' direction, and a way ``move'' on a manifold. Below we show the concrete development of these objects that allow to do develop a second-order trust-region algorithm on manifolds. \changeBMM{The concrete manifold-related ingredients are shown in Table \ref{tab:spaces}, which are based on the developments in \cite{mishra12a}.}

Since the manifold $\mathcal{M}/\sim$ is an abstract space, the elements of its tangent space $T_{[x]} (\mathcal{M}/\sim)$ at $[x]$ also call for a matrix representation in the tangent space $T_x {\mathcal{M}}$ that respects the equivalence relation $\sim$. Equivalently, the matrix representation of $T_{[x]} (\mathcal{M}/\sim)$ should be restricted to the directions in the tangent space $T_{  x }  {\mathcal{M}}$ on the total space $ {\mathcal M}$ at ${  x}$ that do not induce a displacement along the equivalence class $[x]$. This is realized by decomposing $T_{  x}  {\mathcal M}$ into complementary subspaces, the \emph{vertical} and \emph{horizontal} subspaces such that $ \mathcal{V}_{  x}  \oplus \mathcal{H}_{  x} = T_{  x}  {\mathcal M}$. The vertical space $\mathcal{V}_{  x}$ is the tangent space of the equivalence class $[x]$. On the other hand, the horizontal space $\mathcal{H}_{  x}$, which is any complementary subspace to $\mathcal{V}_{  x}$ in $T_x\mathcal{M}$, provides a valid matrix representation of the abstract tangent space $T_{[x]} (\mathcal{M}/\sim)$ \cite[Section~3.5.8]{Absil_2009optimizationonManifolds}. An abstract tangent vector $\xi_{[x]} \in T_{[x]} (\mathcal{M}/\sim)$ at $ [{x}]$ has a unique element in the horizontal space $ {\xi}_{ {x}}\in\mathcal{H}_{ {x}}$ that is called its \emph{horizontal lift}. Our specific choice of the horizontal space is the subspace of $T_x \mathcal{M}$ that is the \emph{orthogonal complement} of $\mathcal{V}_{  x}$ in the sense of a Riemannian metric (an inner product).

%The vertical space $\mathcal{V}_x$ is the linearization of the equivalence class. The horizontal space $\mathcal{H}_{ {x}}$ is complementary to the vertical space $\mathcal{V}_x$ and provides a matrix representation to the abstract tangent space $T_{{[x]}}( {\mathcal{M}}/\sim)$ of the Riemannian quotient manifold. Consequently, tangent vectors on the quotient space are lifted to the horizontal space. {Given $ {\xi}_{  x}$ as the horizontal lift, i.e., matrix representation of a tangent vector $\xi_{[x]}$ belonging to the abstract space $T_{[x]} (\mathcal{M}/\sim)$, $ {R}_{  x}$ maps it onto an element in $\mathcal{M}$.}

\changeBMM{A Riemannian metric or an inner product $ {g}_{  x} : T_x \mathcal{M} \times T_x \mathcal{M}
\rightarrow \mathbb{R}$ at $ {x} \in \mathcal{  M}$ in the total space defines
a Riemannian metric $g_{[x]}:T_{[x]} (\mathcal{M}/\sim) \times T_{[x]} (\mathcal{M}/\sim)
\rightarrow \mathbb{R}$, i.e.,
\begin{equation}\label{eq:metric_quotient}
        g_{[x]}(\xi_{[x]},\eta_{[x]}):= {g}_{ {x}}( {\xi}_{ {x}}, {\eta}_{
{x}}),
\end{equation}
on the quotient manifold $\mathcal M/\sim$, provided that the expression ${g}_{ {x}}( {\xi}_{ {x}}, {\eta}_{ {x}})$ does not depend on a specific representation along the equivalence class $[x]$. Here $\xi_{[x]}$ and $\eta_{[x]}$ are tangent vectors in $T_{[x]} (\mathcal{M}/\sim)$, and $ {\xi}_{ {x}}, {\eta}_{ {x}}$ are their horizontal lifts in $\mathcal{H}_{ {x}}$ at $x$. Equivalently, if $y$ is another element that belongs to $[x]$ and $\xi_y$ and $\eta_y$ are the horizontal lifts of $\xi_{[x]}$ and $\eta_{[x]}$ at $y$, then the metric in (\ref{eq:metric_quotient}) obeys the equality ${g}_{ {x}}( {\xi}_{ {x}}, {\eta}_{
{x}}) =  {g}_{ {y}}( {\xi}_{ {y}}, {\eta}_{
{y}})$. Such a metric is then said to be \emph{invariant} to the equivalence relation $\sim$.}

\changeBMM{In the context of fixed-rank matrices, there exist metrics which are invariant.} A particular invariant Riemannian metric on the total space ${\mathcal{M}}$ that takes into account the symmetry (\ref{eq:symmetry_gh}) imposed by the factorization model and that is well suited to a least-squares objective function \cite{mishra12a} is
\begin{equation}\label{eq:metric_gh}
%\begin{array}{lll}
{g}_{{x}}
 (    {  \xi}_{{x}} ,  {\eta}_{{x}}  )   =  \trace  ((\mat{V}^T\mat{V}){\xi}_{\mat
U}^T {\eta}_{\mat U})   +  \trace ( (\mat{U}^T\mat{U}) {\xi}^T_{\mat V} {\eta}_{\mat
V}  ),
%\end{array}
\end{equation}
where ${x} = (\mat{U}, \mat{V})$ and ${\xi}_{{x}},{\eta}_{{x}} \in T_{{x}}
{\mathcal{M}}$. \changeBM{It should be noted that the tangent space $T_x\mathcal{M}$ has the matrix characterization $\mathbb{R}^{K\times r} \times \mathbb{R}^{K\times r}$, i.e., $\eta_x$ (and similarly $\xi_x$) has the matrix representation $(\eta_{\mat U}, \eta_{\mat V}) \in \mathbb{R}^{K\times r} \times \mathbb{R}^{K\times r}$.} 

\changeBMM{To show that (\ref{eq:metric_gh}) is invariant to the transformation (\ref{eq:symmetry_gh}), we assume that another element $y \in [x]$ has matrix representation $(\mat{U} \mat{M}^{-1}, \mat{V} \mat{M})$ for a non singular square matrix $\mat{M}$. Similarly, we assume that the tangent vector $\eta_{y}$ (similarly $\xi_y$) has matrix representation $(\eta_{{\mat U} \mat{M}^{-1}}, \eta_{{\mat V}\mat{M}^T}) \in \mathbb{R}^{K\times r} \times \mathbb{R}^{K\times r}$. If $\eta_x$ and $\eta_y$ (similarly for $\xi_x$ and $\xi_y$) are the \emph{horizontal lifts} of $\eta_{[x]}$ at $x$ and $y$, respectively. Then, we have $\eta_{{\mat U} \mat{M}^{-1}} = \eta_{{\mat U}}  \mat{M}^{-1}$ and $\eta_{{\mat V}\mat{M}} = \eta_{{\mat V}}\mat{M}^T$ \cite[Example~3.5.4]{Absil_2009optimizationonManifolds}. Similarly for $\xi_y$. A few computations then show that ${g}_{ {x}}( {\xi}_{ {x}}, {\eta}_{
{x}}) =  {g}_{ {y}}( {\xi}_{ {y}}, {\eta}_{
{y}})$, which implies that the metric (\ref{eq:metric_gh}) is invariant to the transformation (\ref{eq:symmetry_gh}) along the equivalence class $[x]$. This implies that we have a unique metric on the quotient space $\mathcal{M}/\sim$.}

\changeBMM{Motivation for the metric (\ref{eq:metric_gh}) comes from the fact that \changeBM{it is induced from a \emph{block diagonal} approximation of the Hessian of a simpler cost function $\| \mat{U}\mat{V}^T - \mat{I}\|_F^2$, which is strictly convex in $\mat{U}$ and $\mat{V}$ individually}. This block diagonal approximation ensures that the cost of computing (\ref{eq:metric_gh}) depends linearly on $K$ and the metric is well suited for least-squares problems.} Similar ideas have also been exploited in \cite{Yuanming_2016LRMCTWC, Mishra_2014r3mc, kasai16a} which show robust performance of Riemannian algorithms for various least-squares problems.

Once the metric (\ref{eq:metric_gh}) is defined on ${\mathcal M}$, \changeBM{the development of the geometric objects required for second-order optimization follow \cite{Absil_2009optimizationonManifolds,mishra12a}}. The matrix characterizations of the tangent space $T_x \mathcal{M}$, vertical space $\mathcal{V}_x$, and horizontal space $\mathcal{H}_x$ are straightforward with the expressions:
\begin{equation}\label{eq:horizontal_space_gh}
\begin{array}{lll}
T_x\mathcal{M} = \mathbb{R}^{K\times r} \times \mathbb{R}^{K\times r}\\
\mathcal{V}_{{x}} =  \{  (   -\mat{U}\mat{\Lambda} , \mat{V}\mat{\Lambda}
^T   ) : \mat{\Lambda} \in \mathbb{R}^{r\times r} \} \\
\mathcal{H}_{{x}}  = \{    ( { \zeta}_{\mat{U}} , {\zeta}_{\mat{U}}    )
:    \mat{U}^T \zeta_{\mat{U}} \mat{V}^T \mat{V}  = \mat{U}^T\mat{U} \zeta_{\mat{V}}^T
\mat{V}     , \\
\qquad \qquad \qquad \qquad  {\zeta}_\mat{U}, {\zeta}_\mat{V}  \in \mathbb{R}^{K
\times r} \}.
 \end{array}
\end{equation}

Apart from the characterization of the horizontal space, we need a linear mapping $\Pi_{{x}}: T_{{x}} {\mathcal{M}} \mapsto \mathcal{H}_{{x}}$ that projects vectors from the tangent space onto the horizontal space. Projecting an element ${\eta}_{{x}} \in T_{{x}} \mathcal{M}$ onto the horizontal space is accomplished with the operator
\begin{equation}\label{eq:projection_gh}
\Pi_{x}({\eta}_{{x}})=  
( {\eta}_{\mat{U}} + \mat{U}\mat{\Lambda} , {\eta}_{\mat{V}} - \mat{V\Lambda}^T
),
\end{equation}
where ${\mat \Lambda} \in \mathbb{R}^{r \times r}$ is uniquely obtained by ensuring that $\Pi_{{x}}({\eta}_{{x}})$ belongs to the horizontal space characterized in (\ref{eq:horizontal_space_gh}). Finally, the expression of $\mat{\Lambda}$ is
\begin{equation*}\label{eq:Lyapunov_gh}
\begin{array}{llll}
& \mat{ U }^T ( {\eta}_{\mat{U}} + \mat{U} {\mat \Lambda}  )  \mat{V}^T \mat{V}
 = \mat{U}^T \mat{U}   (   {\eta}_{\mat{V}} - \mat{V} {\mat \Lambda} ^T
)^T \mat{V} \\

%\Rightarrow &  2 \mat{U}^T\mat{U} \mat{\Lambda}  \mat{V}^T\mat{V} & = & %\mat{U}^T\mat{U}\eta_{\mat{V}}^T\mat{V}
  % - \mat{U}^T\eta_{\mat{U}}  \mat{V}^ T \mat{V} \\

\Rightarrow & \mat{\Lambda}  =   0.5 ( \eta_{\mat{V}}^T\mat{V} (\mat{V}^
T \mat{V})^{-1}  - (\mat{U}^T\mat{U} )^{-1}\mat{U}^T\eta_{\mat{U}} ).
\end{array}
\end{equation*}

\subsection{Gradient and Hessian Computations}\label{sec:gradient_Hessian}
The choice of the metric (\ref{eq:metric_gh}) and of the horizontal space (as the orthogonal complement of $\mathcal{V}_{x}$) turns the quotient manifold $\mathcal{M}/\sim$ into a \emph{Riemannian submersion} of $({\mathcal{M}}, {g})$ \cite[Section~3.6.2]{Absil_2009optimizationonManifolds}. This special construction allows for a convenient matrix representation of the gradient \cite[Section~3.6.2]{Absil_2009optimizationonManifolds} and the Hessian \cite[Proposition~5.3.3]{Absil_2009optimizationonManifolds} on the quotient manifold $\mathcal{M}/\sim$. \changeBMM{Below we show the gradient and Hessian computations for the problem (\ref{eq:manifold_formulation}).}

The Riemannian gradient ${\grad}_{[x]} f$ of $f$ on $\mathcal{M}/\sim$ is uniquely represented by its horizontal lift in ${\mathcal{M}}$ which has the matrix representation
\begin{equation}\label{eq:Riemannian_gradient}
\begin{array}{llll}
{\sf {horizontal\ lift\ of\ }}  { {\grad}_{[x]} f} \\
\qquad \quad = \grad_x  f =  (\frac{\partial f}{\partial
\mat{U}} (\mat{V}^T\mat{V})^{-1},  \frac{\partial f}{\partial
\mat{V}} (\mat{U}^T\mat{U})^{-1}),
\end{array}
\end{equation}
where $\grad_x  f$ is the gradient of $f$ in $\mathcal{M}$ and ${\partial f}/{\partial \mat{U}}$ and ${\partial f}/{\partial \mat{V}}$ are the \emph{partial derivatives} of $f$ with respect to $\mat{U}$ and $\mat{V}$, respectively.

In addition to the Riemannian gradient computation (\ref{eq:Riemannian_gradient}), we also require the directional derivative of the gradient along a search direction. This is captured by a \emph{connection} $\rc _{\xi_x} \eta_x$, which is the \emph{covariant derivative} of vector field $\eta_x$ with respect to the vector field $\xi_x$. The Riemannian connection $\rc_{\xi_{[x]}} \eta_{[x]}$ on the quotient manifold $\mathcal{M}/\sim$ is uniquely represented in terms of the Riemannian connection ${\rc}_{{\xi}_{x}} {\eta}_{x}$ in the total space ${\mathcal{M}}$ \cite[Proposition~5.3.3]{Absil_2009optimizationonManifolds} which is 
\begin{equation} \label{eq:Riemannian_connection}
{\sf{ horizontal\ lift\ of\ }} { {\rc}_{\xi _{[x]}} {\eta _{[x]}}} = \Pi_{{x}}
({\rc}_{{\xi}_{x}} {\eta}_{x}),
\end{equation}
where $\xi_{[x]}$ and $\eta_{[x]}$ are vector fields in $\mathcal{M}/\sim$ and ${\xi}_{x}$ and ${\eta}_{x}$ are their horizontal lifts in ${\mathcal{M}}$. Here $\Pi_{x}(\cdot)$ is the projection operator defined in (\ref{eq:projection_gh}). It now remains to find out the Riemannian connection in the total space ${\mathcal{M}}$. We find the matrix expression by invoking the \emph{Koszul} formula \cite[Theorem~5.3.1]{Absil_2009optimizationonManifolds}. After a routine calculation, the final expression is \cite{mishra12a}
\begin{equation}\label{eq:connection_total_space}
\begin{array}{llll}
{\rc}_{{\xi}_x} {\eta_x}  =  \D {\eta_x}[{\xi_x}] + \left( \mat{A}_{\mat{U}},
\mat{A}_{\mat V} \right), \  {\rm where} \\

\\

\mat{A}_{\mat U}  =  {\eta}_{\mat U}\Sym( {\xi}_{\mat V} ^T \mat{V} )(\mat{V}^T\mat{V})^{-1}
    + {\xi}_{\mat U}\Sym( {\eta}_{\mat V} ^T \mat{V} )(\mat{V}^T\mat{V})^{-1}\\

\quad \qquad - \mat{U}\Sym( {\eta}_{\mat V} ^T {\xi}_{\mat V} ) (\mat{V}^T\mat{V})^{-1}
\\
\mat{A}_{\mat V}  =  {\eta}_{\mat V}\Sym( {\xi}_{\mat U} ^T \mat{U} )(\mat{U}^T\mat{U})^{-1}
    + {\xi}_{\mat V}\Sym( {\eta}_{\mat U} ^T \mat{U} )(\mat{U}^T\mat{U})^{-1}
 \\

\quad \qquad - \mat{V}\Sym( {\eta}_{\mat U} ^T \overline{\xi}_{\mat U} ) (\mat{U}^T\mat{U})^{-1}
\end{array}
\end{equation}
and $ \D {\xi}[{\eta}]$ is the Euclidean directional derivative $ \D {\xi}[{\eta}] : = \lim_{t \rightarrow 0} {({\xi}_{{x} + t {\eta}_{\bar x} } - {\xi}_{x})}/{t}$. $\Sym(\cdot)$ extracts the symmetric part of a square matrix, i.e., $\Sym(\mat{Z}) = ({\mat{Z} + \mat{Z}^T})/{2}$.

The directional derivative of the Riemannian gradient in the direction $\xi_{[x]}$ is given by the \emph{Riemannian Hessian operator} ${\hess_{[x]}} f [\xi _{[x]}]$ which is now directly defined in terms of the Riemannian connection $\rc$. Based on (\ref{eq:Riemannian_connection}) and (\ref{eq:connection_total_space}), the horizontal lift of the Riemannian Hessian in ${\mathcal{M}}/\sim$ has the matrix expression:
\begin{equation}\label{eq:Riemannian_hessian}
{\rm horizontal\ lift\ of\ } {\hess_{[x]}} f [\xi _{[x]}] = \Pi_{{x}}(  {\rc}_{{\xi}_{x}} { {\grad_{ x}} f}   ),
\end{equation}
where  $\xi_{[x]} \in T_{[x]} (\mathcal{M}/\sim)$ and its horizontal lift ${\xi}_{x} \in \mathcal{H}_{{x}} $. $\Pi_x(\cdot)$ is the projection operator defined in (\ref{eq:projection_gh}).

%{For the optimization algorithm in Section \ref{sec:trust_regions} we only require the Riemannian Hessian applied in a particular search direction (instead of the full Hessian) and this is given by $\hess_{[x]} f [\xi _{[x]}]$}. 

%This involves the choice of an \emph{affine connection} $\rc$ on the manifold. The affine connection provides a definition for  the \emph{covariant derivative} of vector field $\eta_x$ with respect to the vector field $\xi_x$, denoted by $\rc _{\xi_x} \eta_x$. Imposing an additional compatibility condition with the metric fixes the
%so-called  \emph{Riemannian} connection which is always unique \cite[Theorem~5.3.1
%and Section~5.2]{Absil_2009optimizationonManifolds}. 

\subsection{Retraction}
An iterative optimization algorithm involves computing a  search direction (e.g., negative gradient) and then ``moving in that direction''. The default option on a Riemannian manifold is to move along geodesics, leading to the definition of the \emph{exponential map}. Because the calculation of the exponential map can be computationally demanding, it is customary in the context of manifold optimization to relax the constraint of moving along geodesics. \changeBM{To this end, we define} {\it retraction} $R_{ x}: \mathcal{H}_{ x}  \rightarrow \mathcal{M}: \xi_x \mapsto R_x(\xi_x)$ \cite[Definition~4.1.1]{Absil_2009optimizationonManifolds}. A natural update on the manifold $\mathcal{M}$ is, therefore, based on the update formula $x_+ = R_x(\xi_x)$, i.e., defined as
\begin{equation}\label{eq:retraction_gh}
\begin{array}{lll}
R_{\mat{U}} ({\xi}_{\mat U}) = \mat{U} + {\xi}_{\mat U} \\
R_{\mat{V}} ({\xi}_{\mat V}) = \mat{V} + {\xi}_{\mat V}, \\
\end{array}
\end{equation}
where ${\xi}_{ x} =(\xi_{\mat U}, \xi_{\mat V}) \in \mathcal{H}_{ x}$ is a search direction and ${x}_+ \in {\mathcal M}$. \changeBM{It translates into the update  $[x_+] = [R_x(\xi_x)]$ on $\mathcal{M}/\sim$.}

\subsection{Riemannian Trust-Region Algorithm}\label{sec:trust_regions}
Analogous to trust-region algorithms in the Euclidean space \cite[Chapter~4]{Wright_2006numericalopt}, trust-region algorithms on a Riemannian quotient manifold with guaranteed superlinear rate convergence and global convergence have been proposed in \cite[Chapter~7]{Absil_2009optimizationonManifolds}. At each iteration we solve the \emph{trust-region sub-problem} on the quotient manifold $\mathcal{M}/\sim$. The trust-region sub-problem is formulated as the minimization of the \emph{locally-quadratic} model of the objective function, say $f : \mathcal{M} \rightarrow \mathbb{R}$ at $x\in \mathcal{M}$,
\begin{equation}\label{eq:TR_subproblem}
\begin{array}{lll}
\mathop{\mini} \limits_{\xi _x \in \mathcal{H}_{x}} \quad & & g_{x} (  \xi _x,   {\grad _x} f ) + \frac{1}{2}  g_{x} (  \xi _x, {\hess_x} f [\xi _x]    ) \\
\subject & & {g}_{x}  ( \xi _x , \xi _x ) \leq \Delta ^2,
\end{array}
\end{equation}
where $\Delta$ is the trust-region radius, $g_x$ is the Riemannian metric (\ref{eq:metric_gh}), and ${\grad_x} f$ and ${\hess_x} f$ are the Riemannian gradient and Riemannian Hessian operations defined in (\ref{eq:Riemannian_gradient}) and (\ref{eq:Riemannian_hessian}), respectively.

Solving the above trust-region sub-problem (\ref{eq:TR_subproblem}) leads to a direction ${\xi}_x$ that minimizes the quadratic model. Depending on whether the decrease of the cost function is sufficient or not, the potential iterate is accepted or rejected. \changeBM{The concrete matrix characterizations} of Riemannian gradient (\ref{eq:Riemannian_gradient}), Riemannian Hessian (\ref{eq:Riemannian_hessian}), projection operator (\ref{eq:projection_gh}), and retraction (\ref{eq:retraction_gh}) allow to use an \emph{off-the-shelf} trust-region implementation on manifolds, e.g., in Manopt \cite{manopt}, which implements \cite[Algorithm $1$]{Absil_2009optimizationonManifolds} that solves the trust-region sub-problem inexactly at every iteration.

\changeBMM{The Riemannian trust-region algorithm is \emph{globally convergent}, i.e., it converges to a critical point starting from any random initialization. The rate of convergence analysis of the algorithm is in \cite[Chapter~7]{Absil_2009optimizationonManifolds}. Theoretically, the algorithm converges to a critical point, but often in practice the convergence is observed to a local minimum. Under certain regularity conditions, the trust-region algorithm shows a \emph{superlinear} rate of convergence locally near a critical point. The recent work \cite{boumal16a} also establishes \emph{worst-case} global rates (i.e., number of iterations required to obtain a fixed accuracy) of convergence over manifolds. In practice, however, we observe better rates.}

\subsection{Computational Complexity}
\changeBMM{The numerical complexity of the algorithm in Algorithm \ref{algua} depends fixed-rank Riemannian optimization algorithm for solving (\ref{eq:regularized_formulation}), (\ref{ls}), and (\ref{eq:refining}) and sorting the diagonal entries of rank-$r$ matrix. The sorting operation depends linearly with $K$ (and logarithmic factors of $K$). The computational cost of the Riemannian algorithm depends on i) the computational cost of the computing the partial derivatives of the objective functions in (\ref{eq:regularized_formulation}), (\ref{ls}), and (\ref{eq:refining}) and ii) the manifold-related operations. The computational cost of the manifold-related ingredients are shown below.}

\begin{enumerate} \itemsep4pt \parskip0pt \parsep0pt
\item \changeBMM{Computation of partial derivatives of the objective functions in (\ref{eq:regularized_formulation}), (\ref{ls}), and (\ref{eq:refining}) with respect to $\mat U$ and $\mat V$:  $O(|\mathcal{V}|r )$.}
\item Computation of Riemannian gradient with the formula (\ref{eq:Riemannian_gradient}): $O(Kr^2 + r^3)$. 
\item Computation of the projection operator (\ref{eq:projection_gh}): $O(Kr^2 + r^3)$. 
\item Computation of retraction $R_{\bar x}$ in (\ref{eq:retraction_gh}): $O(Kr)$. 
\item Computation of Riemannian Hessian with the formulas (\ref{eq:Riemannian_connection}), (\ref{eq:connection_total_space}), and (\ref{eq:Riemannian_hessian}): $O(r^3 + Kr^2)$.
\end{enumerate}
It is clear that all the manifold-related operations are of linear complexity in $K$ and cubic  in $r$. Overall, the cost per iteration of the proposed algorithm in Algorithm \ref{algua} is \emph{linear} with $|\mathcal{V}|$.

%For the case of interest, $r \ll K$, these operations are therefore computationally very efficient. The ingredients that depend on the problem at hand are the evaluation of the cost function $\bar{\phi}$, computation of its partial derivatives and their directional derivatives along a search direction. In the next section, the computations of the partial derivatives and their directional derivatives are presented for the low-rank matrix completion problem.

\section{Simulation Results}
\label{simres}
In this section, we simulate our proposed Riemannian optimization algorithm for user admission control in topological interference management. All the Riemannian algorithms for the rank-constrained optimization problems
(\ref{eq:regularized_formulation}), (\ref{ls}) and (\ref{eq:refining}) are
implemented based on the manifold optimization toolbox Manopt \cite{manopt}.
As the second-order Riemannian trust-region method is robust to the initial points,
all the Riemannian algorithms are initialized randomly and terminated when
either the norm of the Riemannian gradient is below $10^{-6}$ or the number
of iterations exceeds 500.  We set $\|\mathcal{P}_{\Omega}(\bm{X}(\mathcal{S}^{[m]}))-\bm{I}_{|\mathcal{S}^{[m]}|}\|_F/\sqrt{K}=10^{-3}$ in (\ref{ls})
to  check if the affine constraint $\mathcal{P}_{\Omega}(\bm{X}(\mathcal{S}^{[m]}))=\bm{I}_{|\mathcal{S}^{[m]}|}$ is satisfied. 

The proposed algorithm is compared to the following approaches:

\begin{itemize}
\item {\bf{Exhaustive search}}: This is achieved by solving a sequence of problem $\mathscr{F}(\mathcal{S}^{[m]})$ using (\ref{ls}) via exhaustively searching over the set $\mathcal{S}^{[m]}\subseteq\{1,\dots, K\}$. We use the Riemannian trust-region algorithm  in Section {\ref{oqm}} to solve (\ref{ls}). However, the complexity of exhaustive search grows exponentially in the number
of users $K$. 

\item {\bf{Orthogonal scheduling}}: In the conventional orthogonal schemes such as TDMA/FDMA, one can only achieve a symmetric DoF allocation $1/K$ per user \cite{Jafar_TIT2013TIM}. In this case, given the symmetric DoF allocation $1/r$, the number of  admitted users equals $r$.
\end{itemize}

\subsection{Admitted Users versus Achievable DoFs}
\begin{figure}[t]
\center
\includegraphics[scale = 0.48]{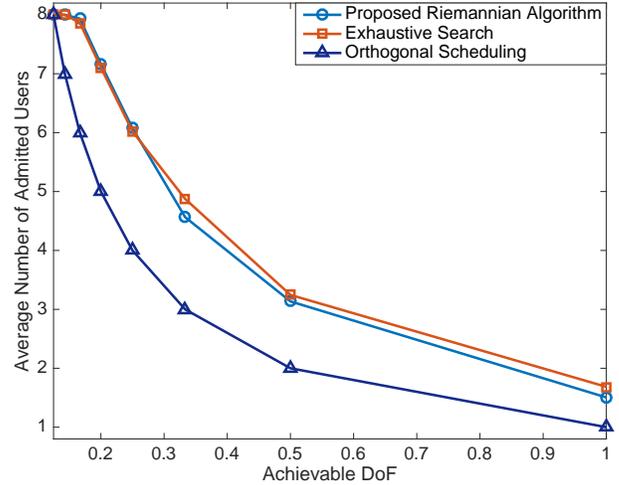}
\caption{Average number of admitted users versus the achievable DoFs with different algorithms.}
\label{dof}
\end{figure}
Consider a 8-user partially connected interference channel with $|\mathcal{V}|=45$ interference channel links. The sets of the connected interference links are generated uniformly at random. The proposed three-stage Riemannian algorithm based user admission approach is compared with the exhaustive search and the orthogonal scheduling approaches.  
We set $\lambda=0.5$, $\rho=0.01$ and $\epsilon=0.01$ in the sparse inducing optimization problem (\ref{eq:regularized_formulation}). Fig. {\ref{dof}} demonstrates  the average number of admitted users with different  symmetric DoF allocations. Each point in the simulations is averaged over 500 randomly generated network topology realizations $\mathcal{V}$. From Fig. {\ref{dof}}, we can see that the proposed three-stage Riemannian algorithm achieves near-optimal performance compared with exhaustive search and significantly outperforms the conventional orthogonal scheduling scheme.

\subsection{Differen Values of the Weighting Parameter $\lambda$} \label{sec:effect_lambda}
\begin{figure}[t]
\center
\includegraphics[scale = 0.48]{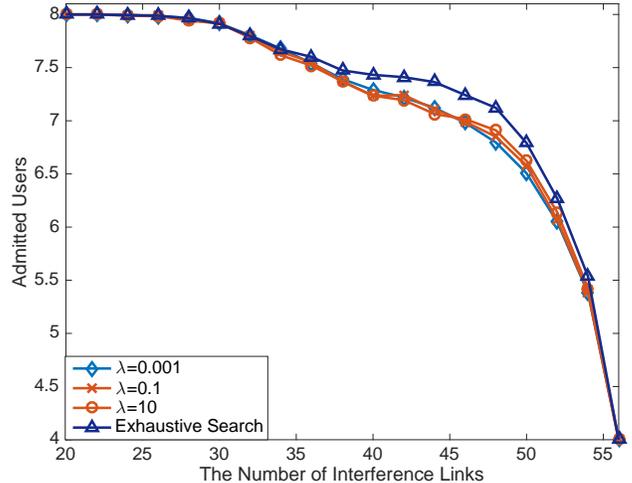}
\caption{Different weighting parameters $\lambda$ in (\ref{eq:regularized_formulation}).}
\label{interference}
\end{figure}

Consider a 8-user partially connected interference channel. The sets of the connected interference links
are generated uniformly at random. We set $\rho=0.01$, $\epsilon=0.01$ and $r=4$ in the sparse inducing
optimization problem (\ref{eq:regularized_formulation}). Fig. {\ref{interference}} shows the average number of admitted users with different values of the weighting parameter $\lambda$ in the regularized smoothed $\ell_1$-norm in (\ref{eq:regularized_formulation}). Each point in the simulations is averaged over 500 randomly
generated network topology realizations $\mathcal{V}$. Fig. {\ref{interference}} demonstrates that parameter $\lambda$ does not affect the induced sparsity pattern in ${\rm{diag}}(\bm{X})$, thereby yielding almost the same number of admitted users. The reason is that the the role of the weighting parameter $\lambda$ in (\ref{eq:regularized_formulation}) only serves to upper bound the objective function. This figure  further indicates that the proposed Riemannian algorithm achieves near optimal performance with the exhaustive search approach and outperforms the orthogonal scheduling scheme with $r=4$, i.e., the number of admitted users is 4.

\section{Conclusions and Discussions}
\label{condis}
This paper presented  a sparse and low-rank optimization framework for user admission control in topological interference management. A Riemannian optimization framework was further developed to solve the non-convex rank-constrained $\ell_0$-norm maximization problem, supported by a novel regularized smoothed $\ell_1$-norm sparsity inducing minimization approach. In particular, by exploiting the quotient manifold of fixed-rank matrices, we presented a  Riemannian trust-region algorithm to find good solutions to the non-convex sparse and low-rank optimization problem. Simulation results illustrated the effectiveness  and near-optimal performance of the proposed algorithms. 

Several future directions of interest are \changeBMM{as follows}:
\begin{itemize}
\item It is desirable but challenging to theoretically establish the fundamental tradeoffs between the sparsity and low-rankness in the sparse and low-rank model $\mathscr{P}$.

\item It is particularly interesting and also important to apply the sparse and low-rank framework to more important problems including the index coding problem \cite{Langberg_TIT2015}, caching networks \cite{Niesen_TIT2014Caching, Tao_2016fundamental}, and distributed computing systems \cite{Ali_CCT16}, thereby investigating the fundamental limits of communication, computation and storage.

\item It is also interesting to apply the Riemannian optimization technique to other important network optimization problems, e.g., blind deconvolution for massive connectivity in Internet-of-Things (IoT) \cite{Romberg_2014blindde} and the hybrid precoding in millimeter wave systems \cite{Letaief_JSTSP2016mmWave}.  
\end{itemize} 

%\newpage
\bibliographystyle{ieeetr}
\bibliography{Reference} % BM
%\bibliography{/Users/Yuanming/Sync/Reference/Reference}

\end{document}